# A NOVEL APPROACH FOR MODELING COMPLEX DEEP FUTURES

Edwin Upchurch[a], Leila Meshkat[b]

[a],[b] California Institute of Technology

[a] etu@caltech.edu, [b] Leila.Meshkat@jpl.nasa.gov

**ABSTRACT**

Many large-scale, complex systems consist of interactions between humans, human-made systems and the environment. The approach developed in this paper is to partition the problem space into two fundamental layers and identify, parameterize and model the main dimensions of each layer and interactions across and in between layers. One layer is the key actors or major organization or human decision makers who influence the state of the world. The other layer includes the domains or fields of knowledge relevant to the problem being addressed. These domains include elements such as the physical earth and its atmosphere, world demography, world economy, level of globalization, and politics. Key parameters for each of the actor types and domains will be extracted and assessed using existing data sources. Novel systems, uncertainty modeling and analysis techniques are combined with advanced computational technologies to determine a spectrum of likely future system states and conduct if-then scenario analyses.

## 1. INTRODUCTION

Senior military and civilian leaders must be enabled to anticipate global trends that may lead to crisis and conflict out to and beyond twenty-five years (so called Deep Futures). Global drivers of change are rapidly evolving and the timelines to respond are increasingly compressed. Proactive response to crisis is even better, effective action before crises manifest is essential to managing change and mitigating or minimizing potential conflict.

There is a rich literature reporting attempts to address the Deep Futures problem, however most of this work addresses pieces of the overall problem. The comprehensive end-to-end world model we feel is necessary for this problem has not yet been built. Our modeling experience is that end-to-end system models are required especially in this case where there are evolving interactions between domains of knowledge, world trends, global actors and their decision mechanisms.

Accomplishing this objective requires creation of analytic and forecasting tools able to help identify and evaluate global drivers of change, their interactions and potential outcomes. This paper articulates the vision, methodology and approach to developing a conceptual model and a solution toolkit. The toolkit is named "Themis".

The technical underpinning of the Themis concept is based on the belief that the risks, mitigations to those risks and opportunities for the actors around the world can be computed sufficiently to add actionable value to users. Themis will be developed as a framework and corresponding methodology which draws from basic mathematical modeling techniques and disciplines and enables orchestration of existing limited scope models into a global model via Themis ontology.

Subject Matter Experts (SME) will remain as the key source of input for developing the conceptual model as well as scenario-specific modifications to the concept. However, Themis will mitigate SME shortcomings by augmenting their knowledge by: (1) actual data feeds that are emerging from internet Big Data and sensor technology breakthroughs; (2) adaptive system technology with SMEs in the loop and (3) high performance computing technology.

It is critical to recognize that there are multiple views (concepts, theories and hypotheses) of how a situation in the world, nation, military, climate, economy, etc. unfolds. Themis will allow simulations to be run with different concepts, theories and hypotheses in order to understand and compare potential future landscapes. In order to provide this information, will include an analysis module, a data repository, a user interface and a scenario generator. The top-level product produced for users of Themis will be an "intervention index" computed from Themis' global assessment of the world situation at a given point in time that can be used as tripwires to trigger in depth analysis and mitigation.

This paper describes development of a Themis conceptual model including how the process must differ from some of the standard ways such models have been attempted in the past. A Themis preliminary concept of operation (CONOPS) is described that leverages Caltech's Jet Propulsion Laboratory's (JPL) Team X approaches to modeling, building and testing highly innovative, one-of-a-kind reliable complex systems. This approach is suitable to Deep Futures modeling which requires coordinating knowledge from a wide variety of global sources including SME's, users, real time and historical data feeds.

## 2. APPROACH

Themis intends to merge expertise and experience in developing kinetic and non-kinetic effects models for the US Army, complex systems modeling and design of innovative space systems and Team X methodology (Meshket, 2006) to develop an evolving comprehensive model of the world that provides users increasingly better responses to key questions. For example: What is the spectrum of driving world trends? What are the first and second order effects of actors' actions? What are the opportunities for specific actor interests? What are the opportunities for an actor's adversaries? to mention a few.

### 2.1. Standard Approaches

In developing the Themis concept current literature on emerging techniques and tools was reviewed. In particular, Waltz (Waltz, 2010) has a good comprehensive treatment of the methodologies, techniques and tools that have emerged due to the fact that "International interventions require unconventional approaches to modeling and analysis". Waltz further points out the reasons why conventional techniques are inadequate: (1) "number and diversity of the participants" and (2) "the effects space spans multiple domains" where there is "a lack of understanding of networked cause-and-effect relationships".

### 2.2. Subject Matter Experts (SME)

The scope of the kind of SMEs needed is greatly increased by Themis. Since Themis must model out 25 or more years, the world must be considered as a system so world SMEs are needed as well as the classical regional and topical experts. World and regional SMEs for the elements of power and therefore the ways that intervention is applied, JIIM+DIMEFIL (Joint Interagency Intergovernmental and Multinational+ Diplomatic Information Military economic Financial Intelligence Law enforcement), are required. World and regional SMEs for modeling dimensions, PMESII+PT (Political Military Economic Social Infrastructure Information+Physical environment Time), are required. There is obvious overlap among these SME sectors of knowledge but the scope still remains large. Finally, world and regional SMEs are required for the Themis domains of: Climate, Demography, Natural Resources, Ideological, Economic, Educational, Health and Health Care, Sociological and Globalization.

SME shortcomings are well known. Themis seeks to mitigate SME shortcomings by augmenting their knowledge by: (1) actual data feeds that are emerging from internet and sensor technology breakthroughs and (2) adaptive system technology with SMEs in the loop.

### 2.3. Alternate Views of the Situation

It is critical to recognize that there are multiple views (concepts, theories and hypotheses) of how a situation in the world, nation, military, climate, economy, etc. operates. By requiring a world model and looking at an extended timeline (25 plus years), Themis has greatly increased the number of distinct and differing theories that may need to be considered in order to understand the scope of future landscapes. For example, there are several global warming and energy resources theories with highly different impacts on potential interventions. There are also a large number of social theories with different significant impacts on future world states. Themis will allow simulations to be run with different concepts, theories and hypotheses in order to understand and compare potential future landscapes.

### 2.4. Key Conceptual Elements

Guided by JIIM+DIMEFIL and PMESII+PT factors with SME guidance Themis will allow the user to identify and develop conceptual representations and relationships for the major elements necessary for a world model. Themis has identified two key conceptual elements: **Domains** and **Actors**.

**Domains** are unique fields of knowledge that have global effects with aggregate potential to drive events that might lead to intervention. The aggregate set of Domains is required to cover the space of key world event drivers. Domains have relationships with other Domains and with Actors.

An initial set of domains that is felt to relate to the types of questions posed to Themis would be determined by SME's. For example, in the Joint Operations Environment (JOE, 2010) Demographics, Globalization, Economics, Energy, Food, Water, Climate Change and Natural Disasters, Pandemics, Cyber, and Space were called out as "trends influencing the world's security". Initial domains are expected to include Climate, Demographical, Resource, Ideological, Educational, Health and Health Care, Sociological and Globalization but Themis will be extensible to additional domains as they are identified or emerge as significant drivers. For example the state of a domain is characterized by the value of its associated parameters as each domain is characterized as a function of these parameters.

**Actors** are decision makers that can influence the state of the system and the actions of the other actors. Actors can be individuals or complex systems such as institutions, organizations, nations, coalitions, multinationals, cartels, etc. Actors have relationships with Domains and other Actors.

## 3. CONCEPTUAL MODEL

Themis includes an analysis module, a data repository, a user interface and a scenario generator. The Themis Repository shall have a well defined ontology and be able to receive data from tools outside of Themis. Once a problem is posed to Themis, it goes through the process of "Problem Formulation". This indicates that it translates the data provided to map to the parameters used to characterize the domains as well as the actors. It further does preliminary analysis to determine the driving parameters in each case and provide high level insight about the system behavior. The next step is to provide this formulated problem to each of the actors and system models for further analysis. Each of these models may use one or more techniques (which may be within the Themis toolbox or may also be outside tools that Themis interfaces with.). When each of these separate layers has been analyzed separately, the information from the analysis is fed into the Themis simulation module. The Themis simulation module will then combine this information and projects it into the future. This information will in turn be fed into the scenario generator which includes modules for uncertainty analysis and aggregation of probabilities in order to propagate uncertainties through the latest system model to determine the likely future scenarios and their associated probabilities. This proposed architecture is shown in Figure 1.

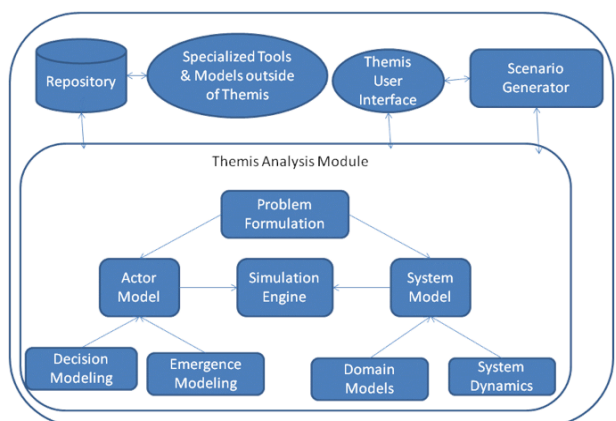

Figure 1: Themis Operations Concept Viewpoint

Figure 2 shows another example viewpoint for the Themis methodology. This viewpoint demonstrates three layers that inter-operate. The lowest level is the system topology. This layer performs the function of representing the system structure. Each of its modules represents a domain of the system or a data source that helps characterize a domain. These domains include the economics, demographics, etc., and different data sources such as the CIA fact-sheet and various reports and databases available. The next layer is the actor decision making layer which performs the function of representing the behavior of key actors. This layer has three modules. One module determines the goals of the actors. Another module the key issues that arise and the interaction between actors as they relate to the resolution of these issues. The third model describes the attitudes of actors with respect to various happenings in the world described in the lowest level. The topmost layer serves to represent the system outcome which is based on the performance of the two first layers. This layer uses relevant information from other layers to conduct analysis and generate the possible scenarios for each part of the world and the likelihoods associated with each of those scenarios. This viewpoint also shows an experiment manager. The experiment manager is responsible for orchestrating the activities of Themis by taking as input the request from the customer and finding a path through Themis that helps to achieve that request.

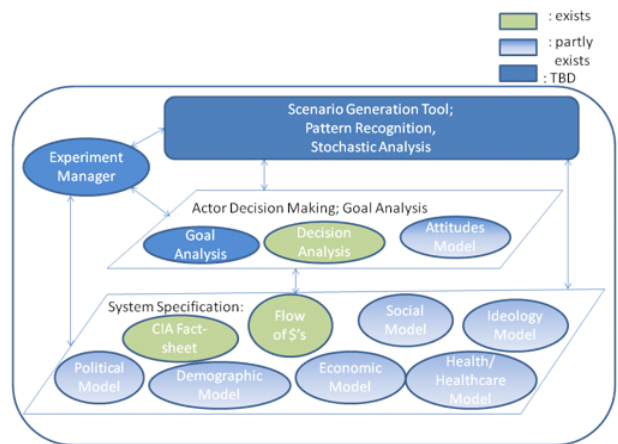

Figure 2: Themis Alternate Viewpoint

Figure 3 shows another perspective on the Themis architecture. Here the Data Analysis layer is shown in tandem with the Themis analysis process. The idea is that the state of the data available is always changing and there is a constant need for analyzing the data and understanding the latest trends and events. This is done outside of Themis and made available to the Themis knowledge/data repository. This repository then makes the data available for the Themis process to run. The Themis user will also create a seed scenario from which the problem definition is created within Themis. The goal and decision analysis then takes place, along with system state analysis. These two are then combined to create the multiple outcomes based on the initial user input and stochastic modeling techniques are used to develop the likelihood of each of these possible outcomes.

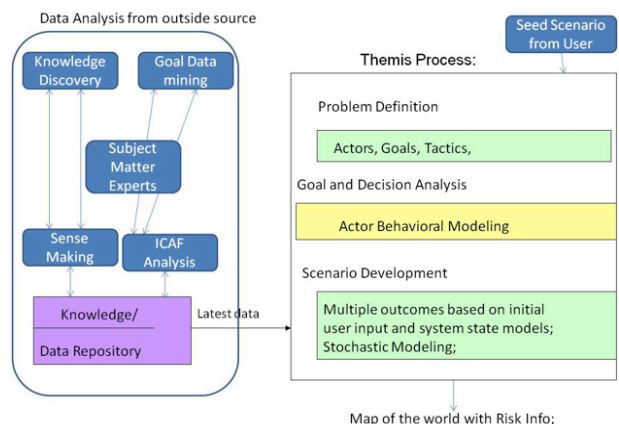

Figure 3: Themis Viewpoint 3

Figure 4 shows a summary of the Themis modeling framework. Themis modeling will be conducted in order to support decisions related to planning, procurement, changes that may occur and force structure. The different types of models that could be used to support Themis can broadly be categorized into actor behavior and domain models and models used for integrating the two. The underpinnings of the actor models are based on social science theories that explain both the actor behaviors as individuals and the social system behavior. The underpinnings of the more physical characteristics of the world, such as climate or demography, are based on science theories. Data used to exercise these models is obtained from the news, or from classified data sources. The history that is relevant to the problem being addressed is obtained from narratives or relevant world scenarios.

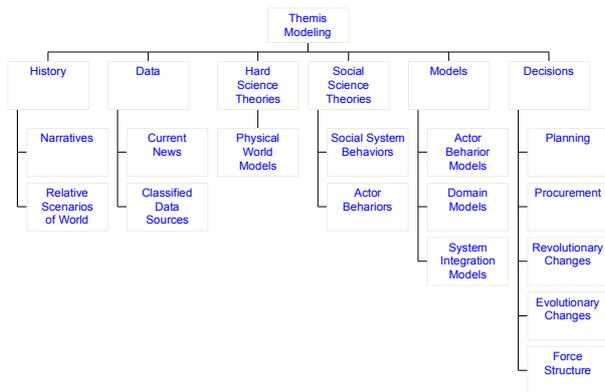

Figure 4: Summary of Themis Modeling

## 4. THEMIS DEVELOPMENT

The first step in development is building a stand-alone "Themis Central" which includes a set of experts using Themis as a tool for conducting multiple types of studies conducting historical use cases examining major trends of the past would be the first set of studies recommended as they help validate the underlying models. The second step is automating some of the functions of the "Themis Central" team so that the interaction between various specialized tools and subject matter experts is semi-automatic. This approach considers Themis to be a member of an interconnected suite of models. These models include other specialized tools and techniques outside of Themis. These tools and techniques are determined by the problem being posed and collaboratively picked by the experts conducting the study.

### 4.1. Central team

The process by which this team operates is shown in Figure 5. This process and team configuration is inspired by the spacecraft conceptual design process used in JPL's concurrent, conceptual design team, TeamX. The customer of the study initially meets with the Themis Central team lead and systems engineer to define the problem of interest. Together the customer and the team leads determine the scope of the study needed to address the problem. It may very well be that an initial study is conducted by the systems engineer and systems modeler with Themis to help the customer scope the problem and determine where they need to focus. Once the problem is well defined, the relevant domain experts are brought in as appropriate. The systems engineer is responsible for designing the study and breaking down the problem into parts to be performed by each of the experts involved. The systems modeler builds the high level systems models within the Themis environment and interfaces with the domain experts as necessary. These two roles may be performed by the same person on small studies. Each of the domain experts work with the systems engineer and system modeler to formulate the problem they need to focus on within their own area of expertise. They then use their expert knowledge as well as specialized tools to analyze this problem and interface with the systems modeler as appropriate.

The transfer of data and information between the various experts and the corresponding models may initially occur manually. As the process becomes well established, the transformations between the different tools and techniques are more and more automated and an underlying data structure and repository is created for the automatic transfer of this information.

Themis will be the tool used by the Systems Modeler in building high level models. The data sources and lower level models that may be required for providing the inputs and insight regarding what experiments to design within Themis to address the questions being posed by the customers are included in the library of tools and techniques used by domain experts.

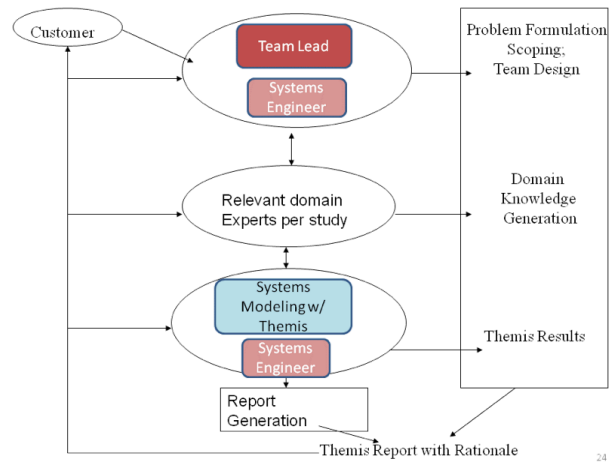

Figure 5: Themis central Team Process

Studies are often conducted iteratively, with the customers and the team leadership reviewing the interim results and identifying the next steps for the study. Once a set of Themis runs are performed, the study lead will synthesize the results to address the questions posed by the customer. Once a set of reasonable answers are created, the results are summarized in a report and provided to the customer. Table 1 summarizes the different suggested roles for a

Themis central team and their corresponding functions and responsibilities.

Table 1: Central Team Roles and Responsibilities

| Role | Function/Responsibilities |
|---|---|
| Customer | • Defines scope of study, duration, results and requirements. |
| Team Lead | • Works with customer to develop study plan.<br>• Works with Systems Engineer to coordinate and plan the team.<br>• Leads team activities and studies.<br>• Synthesizes the results.<br>• Makes executive decisions about direction of the study as it is proceeding. |
| Systems Engineer | • Works with team lead and customer to define the study.<br>• Creates overall design for study. This includes defining the scope, experiments to be conducted and iterations between the Themis model and domain models. |
| Systems Modeler | • Develops Themis models.<br>• Works with Systems Engineer and domain experts to incorporate their information. |
| Domain Expert | • Works with systems modeler and systems engineer to define domain problems.<br>• Uses specialized domain tools to develop models and solutions for domain.<br>• Iterates with systems modeler for Themis design. |

As a problem is posed to Themis, it will use the history related to the problem, the data sets and modeling techniques applicable, current information and the social theory underpinnings that are chosen for the analysis. The models may correspond to the social theories in question. For instance, William Bernstein states in "The Birth of Plenty" (Bernstein, 2010) that there are four factors necessary for a nation to become wealthy: property rights, scientific rationalism, capital markets, and fast and efficient communication and transportation. This could be one model of the state of the economic domain.

The user of Themis or the orchestrator will pick their preferred social theory and corresponding models. For instance, they may pick the history and data related to a specific country and the model that is based on Bernstein's social theory to predict the economic status of the country. The best models for the user to pick would depend on the problem being addressed. There would exist in-depth domain models for each of the domains of interest. Once the right models and knowledge/data bases have been used for their corresponding analysis, it is time to formulate the problem to be solved within the Themis environment. Themis will include high level models of all the relevant domains and is able to combine the results obtained from each specialized model or database to provide insight into the state of the overall system which is the world. The process of using Themis as a member of the orchestra is shown in Figure 6.

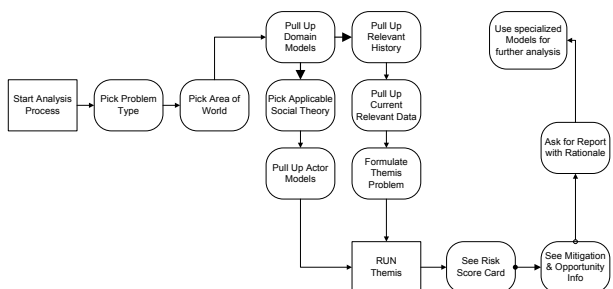

Figure 6: Themis As a Member of the Orchestra

### 4.2. Themis As Orchestrator

Ultimately the intention is that Themis be used as the principal integrator and user interface. Themis will have the ability to automatically combine the data obtained from other models and make the transformation to its repository for Themis modeling. Furthermore, Themis will provide access to other tools and techniques for the user to work with. Note that the key distinction between when Themis is the orchestrator versus when Themis is a member of the orchestra is that the various modeling steps are automatically done by the experiment manager within Themis when Themis is an orchestrator.

Two use cases are shown in Figures 7 and 8. In both of these cases, the user starts by opening Themis and then picks the problem type and relevant domain models, social science theories and other knowledge bases directly from Themis. In the background, there is an intelligent system manager that works with the user to help define the necessary data and models necessary to formulate and solve the problem in Themis. Iterative loops occur within the Themis environment. For instance, in Figure 7, the user decides to change the relevant social science theory and re-run the model.

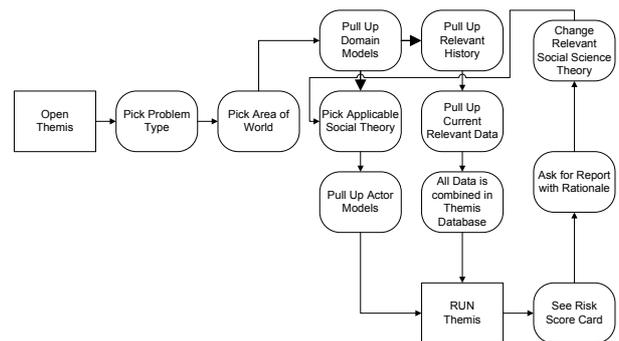

Figure 7: Themis As An Orchestrator (1)

In Figure 8, the user decides to insert additional actors into the system to see how that ripples through.

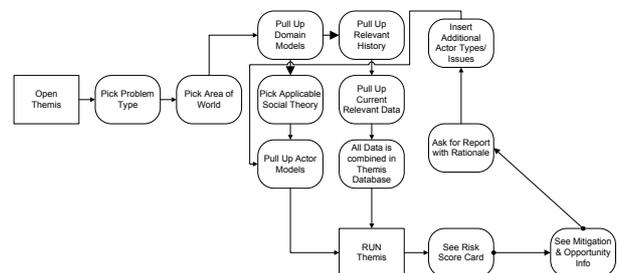

Figure 8: Themis As An Orchestrator (2)

### 5. DEMONSTRATIVE EXAMPLE

For demonstrative purposes, let's consider a simple nominal example. The problem is assessing the likelihood of intervention associated with a country X within the next 25 years. The first step is to characterize that country with parameters associated with its domains.

Either exact or approximate values for each of the parameters of interest over the last 20 years are collected from the existing databases. Figure 9 depicts these key relevant parameters and some of the relationships between them. The complete set of relationships between the various parameters is given in the adjacency matrix shown in Table 2. A "1" in the cell at the intersection of the row and column of the matrix between two parameters indicates that these parameters are related. As it can be see, this is quite a sparse matrix. Therefore, it's very likely that the number of significant variables can be reduced. The next step is to conduct a Principal Component analysis (Draper and Smith, 1998) to determine the key independent variables. Let's assume these key variables include migration, GDP, literacy, religious education, status of women, level of health and potable water. These variables are shown in bold in Figure 9.

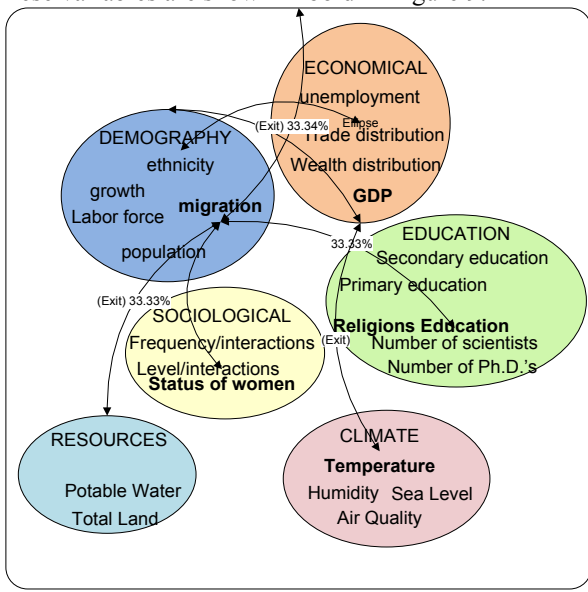

Figure 9: Parameters and Influences for Country X

Further statistical analysis of data associated with these parameters for the last 20 years indicates the type of effect they have on each other. This is shown in table 3. At the intersection of the row "migration" and the column "GDP" there is a "-" sign. This indicates that an increase in the variable "migration" causes a decrease in the variable "GDP". There is a "+" sign at the intersection of the row "status of women" and the column "level of health". This indicates that based on the existing data, an increase in the status of women causes an increase in the level of health. Therefore the key parameters of significance and relationships between them are obtained by analyzing existing data. It's possible to extrapolate this data to assess the state of these key parameters over the next 25 years. But the state of the world depends not only on the trends associated with the domains, but also the behavior of key actors for country X.

Table 2: Adjacency Matrix for Country X

Table 3: Relationship Between Key Variables

|  | migration | GDP | Literacy | Religious Education | Level of health | status of women | Potable water |
|---|---|---|---|---|---|---|---|
| migration | 1 | - | - |  | - | x | + |
| GDP | - | 1 | + | - | + | + | + |
| Literacy | + | + | 1 | x | + | + | x |
| Religious Education | + | - | + | 1 | x | - | x |
| Level of health | - | + | + | x | 1 | x | + |
| status of women | - | + | + | - | + | 1 | + |
| Potable Water | - | + | x | x | + | x | 1 |

Based on existing data, as well as subject matter expertise, the actors associated with that region are classified into three different types A, B and C. Each type is characterized by a set of parameters as well as a main goal. In order to determine the relative influence of each type of actor and hence the most likely state of the system, the goals of each actor are articulated using a linear objective function. The constraints associated with the domains are also articulated and the set of linear equations and constraints is solved via Linear and Goal Programming approaches (Scniederjans, 1995).

These set of equations are given values of the domain parameters associated with the point in time which is of interest. Since the values of the parameters were assessed for the next 25 years, the level of goal attainment of each actor, which in this case corresponds to the is increasing the wealth and population of their respective supporters is obtained by solving the Linear Programming problem.

Given the level of goal attainment for each actor and the basic understanding about the country in question from related data (which includes historical data and expert information), the Themis modeling engine will be able to then generate the scenario which leads to a high risk state and will conduct probabilistic analysis to determine the likelihood for intervention.

Figure 10 illustrates one such scenario. Religious dogmatism causes a reduction in GDP as well as a reduction in the status of women. Water shortage in turn causes disease and migration. Migration causes a reduction in GDP as does the decrease in the status of women. As the migration increases, the average level of education within the society decreases and this in turn causes a reduction in GDP as well. Once the GDP becomes lower than a certain threshold, there is civil unrest that causes the government to lose control and necessitates intervention. Using the available trends and data, Themis will estimate the value of the probabilities associated with the root events and the conditional probabilities associated with the other events to build the Bayesian Belief Network (BBN) associated with this scenario. Solving this BBN using the estimated probability values indicates a 62% chance of the need for US intervention. These values are shown in Figure 11.

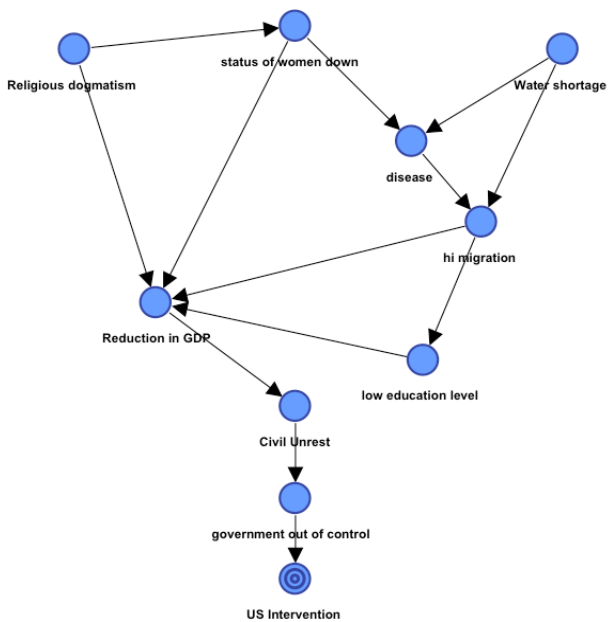

Figure 10: Scenario and Associated BBN

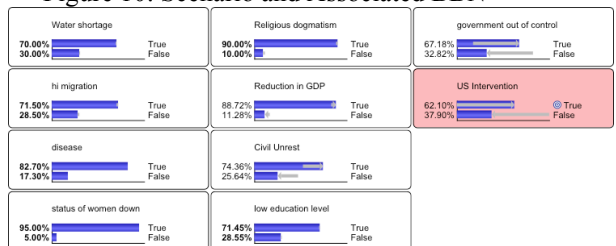

Figure 11: Probability Distribution for the BBN

## 6. CONCLUSION

The key features of this paper include:

- Re-phrasing of the original problem statement into risk and opportunity analysis which can be modeled.
- Summarization/extrapolation of trend analysis relating to the future expectation of the Joint Operational Environment for the Army.
- Adding dimensions to the definition of risk. These dimensions include measures for the direction and acceleration the risk is taking. Seeks to represent the problem space with a few key parameters.
- Partitioning of the problem space into two fundamental layers (actor and domain).
- Preliminary classification of actors and key drivers for their decisions.
- Preliminary assessment of how different "types" of actors emerge and how emergence is interdependent on the state of the domains.
- Defining a team structure based on JPL's experience in engineering of large scale, complex spacecraft with multi-disciplinary teams.